\documentclass[twocolumn,floatfix,showpacs,preprintnumbers,amsmath,amssymb,superscriptaddress]{revtex4}
\usepackage{epsfig}
\usepackage{color}
\usepackage{amsmath}
\usepackage{amssymb}
\usepackage{citesort}
\usepackage{psfrag}
\usepackage{pstricks}
\usepackage{longtable}
\usepackage{dcolumn}
\usepackage{bm}

\newcommand{\ket}[1]{\mathinner{|{#1}\rangle}}

\newcommand{\braket}[2]{\langle #1|#2\rangle}
\def\c#1{\textcolor{red}{#1}}

\providecommand{\abs}[1]{\lvert#1\rvert}
\providecommand{\bigabs}[1]{\bigl\lvert#1\bigr\rvert}

\begin{document}

\title{A study of heuristic guesses for adiabatic quantum computation}

\author{Alejandro Perdomo}
\affiliation{Department of Chemistry and Chemical Biology, Harvard University, 12 Oxford Street, 02138, Cambridge, MA}

\author{Salvador E. Venegas-Andraca}
\affiliation{Quantum Information Processing Group. Tecnol\'{o}gico de Monterrey Campus Estado de M\'{e}xico.
Carretera Lago Gpe. Km 3.5, Atizap\'{a}n de Zaragoza, Edo. M\'{e}xico, M\'{e}xico}.
\affiliation{Department  of Chemistry and Chemical Biology, Harvard University, 12 Oxford Street, 02138, Cambridge, MA}
\email{svenegas@itesm.mx}

\author{Al\'an Aspuru-Guzik}
\affiliation{Department of Chemistry and Chemical Biology, Harvard University, 12 Oxford Street, 02138, Cambridge, MA}
\email{aspuru@chemistry.harvard.edu}

\begin{abstract}
Adiabatic quantum computation (AQC) is a universal model for quantum computation which seeks to transform the initial ground state
of a quantum system into a final ground state encoding the answer to a computational problem. AQC initial Hamiltonians conventionally
have a uniform superposition as ground state. We diverge from this practice by introducing a simple form of heuristics: the ability to start the quantum evolution with a state which is a guess to the solution of the problem.
With this goal in mind, we explain the viability of this approach and the needed modifications to the conventional AQC (CAQC) algorithm.
By performing a numerical study on hard-to-satisfy 6 and 7 bit random instances of the satisfiability problem (3-SAT), we show how
this heuristic approach is possible and we identify that the performance of the particular algorithm proposed is largely determined by
the Hamming distance of the chosen initial guess state with respect to the solution.  Besides the possibility of introducing educated
guesses as initial states, the new strategy allows for the possibility of restarting a failed adiabatic process from
the measured excited state as opposed to restarting from the full superposition of states as in CAQC. The outcome of the measurement can be used as a more refined guess state to restart
the adiabatic evolution. This concatenated restart process is another heuristic that the CAQC strategy cannot capture.
\end{abstract}

\pacs{03.67.Lx, 03.67.Ac, 03.65.-w }

\maketitle

\section{Introduction}\label{sec:intro}

Adiabatic quantum computation (AQC) \cite{Farhi2000} is a promising paradigm of quantum computation
because of its robustness \cite{childs01,lidar08}, and its intuitive mapping from NP-complete and NP-hard problems to potentially realizable Hamiltonians \cite{Farhi2000,Farhi2001,hogg03,young08}. Adiabatic quantum computing is attractive because relevant optimization problems such as lattice models for protein folding can be readily formulated \cite{perdomo08}.

AQC algorithms involve the specification of a time-dependent Hamiltonian,
\begin{equation}
\hat{H}(t) = \hat{h}_i(t) + \hat{h}_{driving}(t) + \hat{h}_f(t),
\end{equation}
This Hamiltonian has three important functions: (1) The initial Hamiltonian, $\hat{H}_i \equiv \hat{H}(0)$, encodes a ground state that is easy to prepare and that is used as the initial state
for the quantum evolution. (2) The driving Hamiltonian, $\hat{h}_{driving}(t)$, is responsible for mediating the transformation of the initial ground state to any of other state. (3) The final Hamiltonian,
$\hat{H}_{f} \equiv \hat{H}(\tau)$, is problem dependent and its ground state encodes the solution, $\ket{\psi_{solution}}$,
to the computational problem. In the ideal case of a process being fully adiabatic, evolution under $\hat{H}(t)$ will keep
the quantum state, $\ket{\psi(t)}$, in the ground state of $\hat{H}(t)$ throughout $0 < t < \tau$. If this condition is met, the final state
at $t=\tau$ should coincide with the ground state of the final Hamiltonian, $\hat{H}_{f}$, i.e., $\ket{\psi(\tau)} = \ket{\psi_{solution}}$,
if the process is adiabatic. The measurement at $t=\tau$ will provide the solution to the computational problem.

Since the original proposal by Farhi et al \cite{Farhi2000}, a significant amount of progress has been made towards the design
of final Hamiltonians for different computationally intractable problems such as NP-complete problems. For example,
in this paper we provide a detailed description of the construction of final Hamiltonian for the 3-SAT problem. In previous work,
we described the respective construction of the final Hamiltonian for an NP-hard problem of interest in biology, the protein folding problem, which consists of finding the minimum energy configuration of a chain of interacting amino acids in a lattice \cite{perdomo08}. Several choices for $\hat{h}_{driving}$ have been suggested. Farhi et al \cite{Farhi2000} proposed what we call in this work the conventional form used for $\hat{h}_{driving}(t)$: a linear ramp,
\begin{equation}
\hat{h}_{driving}(t) = (1-t/\tau) \hat{H}_{driving},
\end{equation}
In Farhi \textit{et al}'s scheme the term $\hat{h}_i(t)$ can be defined as $\hat{H}_i \equiv \hat{h}_{driving}(t=0) = \hat{H}_{driving}$, namely, the term $\hat{H}_{driving}$ serves as the initial Hamiltonian $\hat{H}_i$ when its intensity is completely turned on at $t=0$, and drives the quantum evolution until it is completely turned off at $t=\tau$.  The final Hamiltonian is turned on with the functional form
$\hat{h}_f(t) = (t/\tau) \hat{H}_f$. Scheduling the adiabatic evolution with this linear interpolation is not compulsory, thus
different proposals have been studied such as the use of non-linear time-profile for auxiliary Hamiltonians \cite{farhi_quantum_2002} and optimal geodesic paths \cite{rezakhani_quantum_2009}.

The question of whether AQC, or in general quantum computers, can be used for efficiently solving NP-complete problems is
a difficult open question. Lessons that have been learned \cite{farhi_to_2008} include that a poor choice
of the initial Hamiltonian such as the one-dimensional projector as selected in Refs. \cite{roland_quantum_2002} and \cite{znidaric_exponential_2006},
will lead inefficient AQC algorithms. Therefore, it is important to consider different strategies which might allow
an escape from bottlenecks or trap states which might limit the use of AQC \cite{amin2008}.

When attempting to tackle combinatorial and optimization problems with classical computers, a common approach to cope with
intractability and NP-complete problems \cite{garey79,sipser05}, is to employ heuristic algorithms as an alternative
to exhaustive search which scales exponentially with system size. In quantum computation, Hogg \cite{hogg_highly_1998,hogg_quantum_2000}
introduced heuristic techniques in quantum algorithms. He showed that using information about the structure
of the problem as a heuristic guide can be used to enhance the performance of quantum search compared to the scheme proposed by Grover \cite{grover_quantum_1997}. Hogg's proposal was suggested for the gate model for quantum computation, but to our knowledge, there are no studies of heuristics in AQC. The purpose of this paper is to examine the following questions: How can we incorporate heuristics in AQC?
Is there any advantage by doing so? What are the modifications to CAQC proposal from an algorithmic and experimental point of view?

\paragraph{Initial state selection.} The simplest form of heuristics we could think of is to start the quantum evolution from a quantum state which is a guess to the solution, but this possibility is not available in any of the proposals for AQC. Physical intuition as well as constraints within the problem can be used to make an educated guess. To illustrate our idea, let's use a lattice model for protein folding \cite{perdomo08} as an example. For this model, an educated guess for the initial state would be to choose a bit string which encodes an initial position for the amino acids in the spatial lattice such that no two amino acids are on top of each other, and that they are connected according to the sequence defining the protein to be folded. Conversely, CAQC would begin the computation with a quantum superposition of all possible states of
the computational basis. This choice of initial state contains absurd configurations like all amino acids on top of each other,
or assignments referring to configurations of amino acids which are fully disconnected or not properly linked according to the protein sequence. We conjecture here that the presence of these ``non-sensical'' states might act as trap states, making a smooth transition towards the desired final ground state more difficult and therefore increasing the time required for an adiabatic evolution. In other fields of computer science, physics and chemistry, one might also use classical methods or a mean field approach to find approximate solutions to be employed as educated guesses. For example, in the context of quantum simulation, a Hartree-Fock solution may be used as the initial state for an adiabatic preparation of an exact molecular wave function \cite{Aspuru-Guzik2005,ward_preparation_2009,wang_quantum_2008,wang_efficient_2009}.

In quantum mechanics, preparing a desired state is not a trivial task \cite{kohen_quantum_1993,aharonov_adiabatic_2003}
and it is not always possible to deterministically prepare a state from a superposition state by measurement. For NP-complete classical problems, like the 3-satisfiability problem (3-SAT) \cite{garey79,sipser05} studied here, the simplest guess for the initial state is that of choosing one of the possible assignments from the solution space.
We present a strategy which allows to design experimentally-realizable initial Hamiltonians whose ground state corresponds to
the desired guess as the initial quantum state. This is essential for studying the importance of choosing a guess state instead of
having full superposition of states as in CAQC \cite{Farhi2000}. In an algorithm like
the one proposed here, both the initial guess and the final solution will be states of the computational basis. Therefore, the overlap between them is zero, unless one has guessed the right solution or the guess is included in the subspace of solutions, which is very unlikely.
Regardless of this counterintuitive choice, $\braket{\psi_{guess}}{\psi_{solution}}=0$, we show that using this kind of heuristics can be of advantage. Even in the case of choosing the initial state by random guessing there is potential for outperforming CAQC.

\paragraph{Restarting the evolution.}  Notice that the method described here not only allows to begin the evolution from a guess state, but also allows for the possibility of restarting a failed AQC calculation from the measured state. This state can be used as a refined guess to restart the adiabatic evolution. An adiabatic processes is an idealized concept because real experiments have to be run in finite time
and therefore there will always be probability of measuring a non-desired excited state which does not encode the solution to
the computational task. The possibility of restarting the quantum evolution using the measured state as a guess is a feature
which is not available in any of the AQC proposals to date, to our knowledge.

The incorporation of heuristics in AQC essentially involves two modifications to CAQC.
We address both changes\c{; }the first modification involves the design of initial Hamiltonians
for arbitrary guess states. The second modifications involves the change of the time profile of the driving Hamiltonian from the linear ramp to a non-linear time dependence with a ``sombrero-like" time profile (see Fig.~\ref{fig:tradit-vs-sombrero}).
Note that this second change is \textbf{not} the main point of the paper since non-linear paths had been proposed before \cite{farhi_quantum_2002}. Also, it is not our purpose to explore what is the optimal selection for the driving term. It must be emphasized that the ``sombrero-like" time profile is an essential feature needed if one is interested in the kind of heuristics we describe here,
but this is not the case in the conventional way of doing AQC where it was used for auxiliary Hamiltonian terms \cite{farhi_quantum_2002}. Because of
this distinctive feature and with the purpose of differentiating our heuristic strategy proposed with CAQC, we will refer to our method as Sombrero Adiabatic Quantum Computation (SAQC). The name should be associated with the algorithmic strategy (selection of initial guess, design of initial Hamiltonian and sombrero-like profile for the driving Hamiltonian)
which aims to incorporate heuristics in AQC, not only to the use of non-linear paths in AQC.

The paper is divided as follows: in Section~\ref{sec:caqc}, we review the CAQC approach. Section~\ref{sec:saqc} introduces the basic elements of the new implementation, SAQC. Finally, in
Section~\ref{sec:numresults}, we present numerical calculations comparing the performance of both the CAQC and the SAQC algorithms based on the minimum gap, $g_{min}$, of their respective time-dependent
Hamiltonians driving their corresponding time evolutions.

\section{Conventional adiabatic quantum computation (CAQC)}\label{sec:caqc}

The goal of AQC algorithms is that of transforming an initial ground state $\ket{\psi(0)}$
into a final ground state $\ket{\psi(\tau)}$, which encodes the answer to the problem. This is achieved by evolving
the corresponding physical system according to the Schr\"{o}dinger equation with a time-dependent
Hamiltonian $\hat H(t)$. The AQC algorithm relies on the quantum adiabatic theorem
\cite{Messiah,marzlin_inconsistency_2004,tong_sufficiency_2007,wei2007,zhao2008,amin_inconsistency_2008,mackenzie_validity_2007,ambainis_elementary_2004,jansen_bounds_2007,wu_adiabatic_2007,chen_invariant_2007}, which states that if the quantum evolution is initialized with the ground state of the initial Hamiltonian, the time propagation of this quantum state will remain very close
to the instantaneous ground state $\ket{\psi_{g} (t)}$ for all $t \in [0, \tau]$, whenever $\hat H(t)$
varies slowly throughout the propagation time $t \in [0, \tau]$. This holds under the assumption that the ground state manifold does not cross the energy levels which lead to excited states of the final Hamiltonian. Here, we denote by ground state manifold the first $m$ curves associated with the lowest eigenvalue of the time-dependent Hamiltonian for $t \in [0, \tau]$, where $m$ is the degeneracy of the final Hamiltonian ground state. An example of $m=2$, is shown in Fig. 5 of Ref.~\cite{perdomo08}.

Conventionally the adiabatic evolution path is the linear sweep of $s \in [0,1]$, where $s = t/\tau$:
\begin{equation} \label{eq:conventional-aqc}
H(s) = (1-s) H_{transverse} + s H_{f}.
\end{equation}
$\hat{H}_{transverse}$ (see Eq.~\ref{eq:hamilt_perturbation} below) is usually chosen such that
its ground state is a uniform superposition of all possible $2^{n}$ computational basis vectors, for the case of an $n-$qubit system. Here, we choose the spin states \{$\ket{q_{i} = 0},
\ket{q = 1}$\}, which are the eigenvectors of $\hat{\sigma}_{i}^{z}$ with eigenvalues +1 and -1, respectively, as the basis vectors. Then the initial ground state
is $\ket{\psi_{g} (0)} = \frac{1}{\sqrt{2^{n}}} \sum_{q_{i} \in \{0,1\}} \ket{q_{n}} \ket{q_{n-1}} \cdots \ket{q_2} \ket{q_1}$. Such an
initial ground state is usually assumed to be easy to prepare, for example, by imposing a global transverse field. Since each state encodes a possible solution, this initial state assigns equal probability to all possible solutions to the computational problem.

\section{Sombrero adiabatic quantum computation (SAQC)}\label{sec:saqc}

For SAQC, the time-dependent Hamiltonian can be written as:
\begin{equation}\label{eq:sombrero-aqc}
\hat{H}_{sombrero} = (1-s) \hat{H}_{i} + \text{hat(s)} \hat{H}_{driving} + s \hat{H}_{f}.
\end{equation}

\begin{figure*}[p]
\begin{center}
 \includegraphics[width=0.9\textwidth]{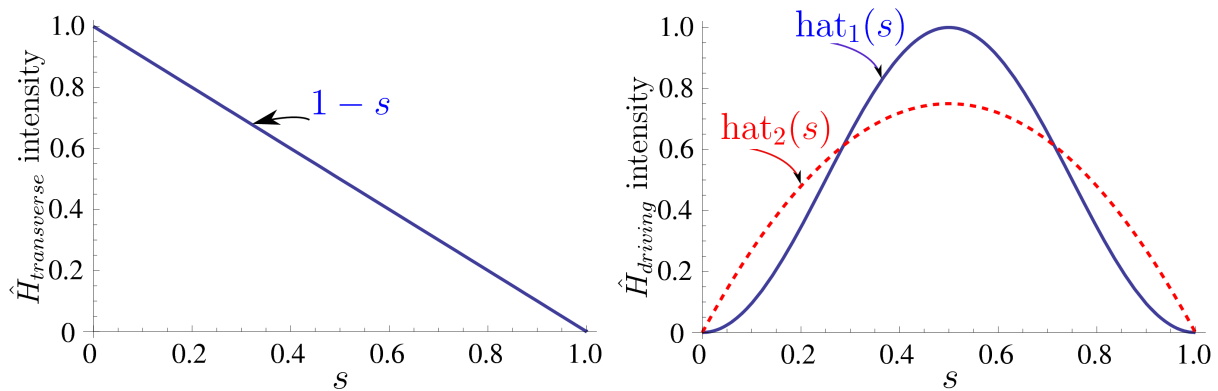}
\end{center}
\caption{\label{fig:tradit-vs-sombrero}(Color online) The main idea behind Sombrero Adiabatic Quantum Computation (SAQC) is to introduce heuristics in AQC, and having the possibility of restarting a failed AQC run from the measured excited state. In order to prepare an arbitrary state from any of the $2^N$ possible basis states from the computational basis of the $N$ qubit system, we propose an initial Hamiltonian, $\hat{H}_i$ (see Eq.~\ref{h-initial}), in such a way that the desired initial guess state is the non-degenerate ground state of the designed initial Hamiltonian. The initial Hamiltonian is diagonal in the computational basis, and so is the final Hamiltonian for the case of classical problems such as the NP-complete problems, e.g., random 3-SAT. Since both, the initial and final Hamiltonians are diagonal, connecting them via a linear ramp as is usually done in CAQC (see left panel) will not lead the quantum evolution towards finding the ground state of the final Hamiltonian. To maintain the initial Hamiltonian uniquely and fully turned on at the beginning, $t=0$, and the final Hamiltonian uniquely and fully turned on at the end of the computation, $t= \tau$, we introduce a driving Hamiltonian whose time profile intensity has a ``sombrero-like" shape (see right panel) is such a way that it only acts during $0< t < \tau$. Two examples of functions with this functional form are presented, where hat$_1(s)=\sin^2 (\pi s)$ and hat$_2(s)=s(1-s)$. A desired feature of our algorithmic strategy is the possibility of introducing heuristics, and not that of introducing non-linear paths. The latter has been proposed previous publications \cite{farhi_quantum_2002,andrecut2004,siu2007}, but here is employed as a consequence of the algorithmic strategy.}

\end{figure*}

We want the non-degenerate ground state of the initial Hamiltonian $\hat{H}_{i}$ to encode a guess to the solution, and
the driving term, $\hat H_{driving}$, to couple the states in the computational basis.
The function $\text{hat}(s)$ is zero at the beginning and end of the adiabatic path; therefore $\hat H_{driving}$ acts only in the range $\text{s} \in(0,1)$ in a ``sombrero-like'' time dependence (see Fig.~\ref{fig:tradit-vs-sombrero}), which allows $\hat{H}_i$ ($\hat{H}_f$) to be fully turned on at the beginning (end) of the computation.

\subsection{Design of the initial Hamiltonian for the guess state}\label{subsec:h-init}

As preparing an arbitrary initial non-degenerate ground state for adiabatic evolution is not
a trivial task, we focus on easy to prepare initial guesses that consist of one of the states in
the computational basis. The strategy proposed builds initial Hamiltonians such that the initial guess
corresponds to the non-degenerate ground state of the initial Hamiltonian, as it is required by AQC.
Additionally, this ground state would be non-degenerate.

Let us denote the states of the computational basis of an $N$ qubit system as $\ket{q_N} \ket{q_{N-1}} \cdots \ket{q_1} \equiv \ket{q_N  \cdots q_1}$ where $q_n \in \{0,1\}$. The proposed initial Hamiltonian, whose ground state corresponds to an arbitrary initial guess state of the form $\ket{x_N  \cdots x_1}$, can be written as

\begin{equation}\label{h-initial}
\hat{H}_{i} = \sum_{n=1}^{N} \left(x_n \hat{I} + \hat{q}_n (1-2 x_n)\right) = \sum_{n=1}^{N} \hat{h}_{x_n},
\end{equation}


where each $x_n$ is a boolean variable, $x_n \in \{0,1\}$, while $\hat{q} \equiv \frac{1}{2} (\hat{I} - \hat{\sigma}^{z})$ is a quantum operator acting on
the $n$-th qubit of the multipartite Hilbert space $\mathcal{H}_{N} \otimes \mathcal{H}_{N - 1} \otimes \cdots \otimes \mathcal{H}_n \otimes \cdots \otimes \mathcal{H}_1$. The operator $\hat q_n$ is given by

\begin{equation}\label{operator-qn}
\hat q_{n} = \hat{I}_N \otimes \hat{I}_{N-1} \otimes \cdots \otimes (\hat q)_n \otimes \cdots \otimes \hat{I}_1,
\end{equation}
where $\hat q$ is placed in the $n$th position and the identity operators act on the rest of the Hilbert space.

The states constituting the computational basis, $\ket{0}$ and $\ket{1}$, are eigenvectors of $\hat{\sigma}_{z}$ with eigenvalues $+1$ and $-1$,
and therefore they are also eigenstates of the operator $\hat q$ with eigenvalues 0 and 1 respectively. The logic behind the initial Hamiltonian
in Eq.~\ref{h-initial} then is clear: if $x_n =0$, then $\hat{h}_{x_n=0} = \hat{q}_n$ but in the case of $x_n =1$,
then $\hat{h}_{x_n=1} = \hat{I} - \hat{q}_n$.

As an example, suppose one has a four qubit system, and one wishes to initialize the adiabatic computation
with the state $\ket{x_4 = 1, x_3 = 0, x_2 = 1, x_1 = 0} \equiv \ket{1010}$ which one may choose either randomly
or as an educated guess to the solution. According to Eq.~\ref{h-initial}, the initial Hamiltonian for
the $\ket{1010}$ guess state should be constructed as

\begin{equation}
\label{h-initial_example}
\begin{split}
\hat{H}_{i} &= \hat{h}_{x_4}+\hat{h}_{x_3}+\hat{h}_{x_2}+\hat{h}_{x_1} \\&
= (\hat{I} - \hat{q}_4)+\hat{q}_3 + (\hat{I} - \hat{q}_2) + \hat{q}_1,
\end{split}
\end{equation}

and, clearly

\begin{equation}\label{h-initial_example_apply_to_qubits}
\begin{split}
\hat{H}_{i} \ket{1010} &= \left((\hat{I} - \hat{q}_4)+\hat{q}_3 + (\hat{I} - \hat{q}_2) + \hat{q}_1\right)\ket{1010} \\& 
= 0 \ket{1010}.
\end{split}
\end{equation}

In general, the $2^N$ states of the computational basis are all eigenstates of $\hat{H}_{i}$, and it can be easily verified that the spectrum of $\hat{H}_{i}$ are energies contained in $\{0,\cdots,N\}$. As required, the ground state is also nondegenerate. The other states will have an eigenenergy which equals their Hamming distance to the ground state of the initial Hamiltonian.

\subsection{Driving Hamiltonian}\label{subsec:h-driv}

The encoding of an educated or a random guess into $\hat H_i$ (Eq. \ref{h-initial}) makes both $\hat H_i$ and $\hat H_f$ (Eq. \ref{eq:sombrero-aqc})
diagonal in the computational basis. Therefore, connecting $\hat H_i$ and $\hat H_f$ with a linear ramp (as in Eq. \ref{eq:conventional-aqc}),
namely, omitting the operator $\hat{H}_{driving}$ in the quantum evolution, would yield zero probability of obtaining the state that encodes the unknown solution to the problem starting from the initial guess state. To avoid such a situation, $\hat H_{driving}$ must introduce non-diagonal terms in $\hat{H}_{sombrero}$ (see Eq.~\ref{eq:sombrero-aqc}) that allows the initial state to transform from any arbitrary guess into the solution.


In order to make a fair comparison between CAQC and SAQC (see Eq.~\ref{eq:conventional-aqc} and Eq.~\ref{eq:sombrero-aqc}), we set
\begin{equation}\label{eq:hamilt_perturbation}
 \hat{H}_{driving} = \hat{H}_{transverse} = \delta \sum_{n=1}^N \hat{q}_{n}^{x},
\end{equation}
in Eq.~\ref{eq:sombrero-aqc}, where $\hat q_{n}^{x}$ stands for the quantum operator
$\hat{q}^{x}$ acting on the $n$th qubit of the multipartite Hilbert space
$\mathcal{H}_{N} \otimes \mathcal{H}_{N - 1} \otimes \cdots \otimes \mathcal{H}_n \otimes \cdots \otimes \mathcal{H}_1$.
The operator $\hat{q}_{n}^{x}$ is given by $\hat{I}_{N} \otimes \hat{I}_{N-1} \otimes \cdots \otimes (\hat{q}^{x})_n \otimes \cdots \otimes \hat{I}_1$,
where the operator $\hat{q}^{x}\equiv \frac{1}{2} (\hat{I} - \hat{\sigma}^{x})$ has been placed in the $n$th position,
and the $\hat{I}_{i}$'s are identity operators.
From a physical point of view, the Hamiltonians $\hat{H}_{driving}$ and  $\hat{H}_{transverse}$
can be related to a transverse magnetic field. The intensity of these Hamiltonians is tuned by varying the $\delta$
parameter. If we set $\delta$ to be the same for both adiabatic algorithms, all the dependence of
the transverse field intensity lies on functions $(1-s)$ in Eq.~\ref{eq:conventional-aqc}
and $\text{hat}(s)$ in Eq.~\ref{eq:sombrero-aqc}. A reasonable requirement
for a fair comparison between CAQC and SAQC is that they both provide the same average
intensity of the transverse magnetic field in $s \in [0,1]$. 
A choice of $\text{hat}(s)$
with the same average $\int_{0}^{1} \text{hat}(s) ds= \int_{0}^{1}(1-s) ds  = 1/2$, is $\text{hat}(s) = 3s(1-s)$.

Even though nonlinear evolutions have been proposed in previous articles \cite{farhi_quantum_2002,andrecut2004,siu2007},
our hat$(s)$ function can be as simple or as complicated as desired, as long as $\text{hat}(0)=\text{hat}(1)=0$
is fulfilled. There is plenty of room to optimize the performance by choosing a more convenient hat$(s)$ for
the adiabatic evolution, we emphasize that the additional advantage of SAQC is the possibility of choosing an initial guess. In the next section we present some results obtained based on one of
the simple nonlinear function $\text{hat}(s) = 3s(1-s)$ and discuss the performance of both CAQC and SAQC for random 3-SAT instances.

\section{Hamiltonians for 3-SAT, numerical calculations and discussion}\label{sec:numresults}

In order to provide a proof of concept for SAQC and to test the potential usefulness
of both random and educated guesses in adiabatic evolution, we performed a numerical study on hard-to-satisfy
6- and 7-variable instances of the 3-SAT problem and compared our results with the CAQC approach.
Let us now provide a succinct introduction to the 3-SAT problem as well as to
briefly discuss its relevance in the fields of theoretical and applied computer science.

\subsection{Construction of final Hamiltonians for satisfiability problems and design of numerical calculations}

{\bf The K-SAT Problem}. Let $A=\{e_1, e_2, \ldots, e_n, \bar{e}_1, \bar{e}_2, \ldots, \bar{e}_n \}$
be a set of Boolean variables $E=\{e_i \}$ and their negations $\bar{E}=\{\bar{e}_i \}$. Let us now construct a logical proposition $P$,
defined as $P = \bigwedge_i [(\bigvee_{j=1}^k a_j)] = \bigwedge_i C_i$, where $a_j \in A$, i.e. P is a conjunction of clauses $C_i$ over the set $A$, where each clause consists
of the disjunction of $k$ literals.
Proposition $P$ is a K-SAT instance and the solution of the K-SAT problem, for instance $P$, consists
of finding a set of values for those binary variables upon which $P$ has been built (i.e. a bitstring), so that
replacement of such binary variables for their corresponding binary values makes $P=1$, namely, proposition
$P$ is satisfied. 3-SAT is a particular case of K-SAT for K=3.

For example, let us examine the following instance of the 3-SAT problem.
Let  $E=\{x_1, x_2, x_3, x_4, x_5, x_6 \}$ be a set of binary variables, and therefore the set of literals is $A=E\cup\bar{E}= \{x_1, x_2, \ldots, x_6, \bar{x}_1, \bar{x}_2, \ldots, \bar{x}_6 \}$. Consider a 3-SAT instance specified by the proposition,

$$
\begin{array}{lll}
P &=&  (\bar{x_1} \vee \bar{x_4} \vee \bar{x_5})  \wedge  (\bar{x_2} \vee \bar{x_3} \vee \bar{x_4}) \wedge (x_1 \vee x_2 \vee \bar{x_5})              \wedge   \\
  & &  (x_3 \vee x_4 \vee x_5)                   \wedge (x_4 \vee x_5 \vee \bar{x_6})              \wedge  (\bar{x_1} \vee \bar{x_3} \vee \bar{x_5}) \wedge  \\
  & &  (x_1 \vee \bar{x_2} \vee \bar{x_5})        \wedge  (x_2 \vee \bar{x_3} \vee \bar{x_6})       \wedge (\bar{x_1} \vee \bar{x_2} \vee \bar{x_6})  \wedge   \\
  & &  (x_3 \vee \bar{x_5} \vee \bar{x_6})       \wedge (\bar{x_1} \vee \bar{x_2} \vee \bar{x_4})  \wedge  (x_2 \vee x_3 \vee \bar{x_4})             \wedge    \\
  & &  (x_2 \vee x_5 \vee \bar{x_6})              \wedge (x_2 \vee \bar{x_3} \vee \bar{x_5})       \wedge (\bar{x_2} \vee \bar{x_3} \vee \bar{x_4})  \wedge                    \\
  & &  (x_2 \vee x_3 \vee x_6) \wedge (\bar{x_1} \vee \bar{x_2} \vee \bar{x_3})  \wedge  (\bar{x_1} \vee \bar{x_4} \vee \bar{x_5}) \wedge  \\
  & &  (\bar{x_3} \vee \bar{x_4} \vee \bar{x_6})  \wedge  (\bar{x_4} \vee \bar{x_5} \vee x_6)       \wedge (\bar{x_2} \vee x_3 \vee \bar{x_6})        \wedge   \\
  & & (x_2 \vee x_5 \vee x_6)                   \wedge (x_3 \vee x_5 \vee \bar{x_6})              \wedge  (\bar{x_1} \vee x_3 \vee \bar{x_6})       \wedge \\
  & & (x_3 \vee \bar{x_5} \vee x_6)              \wedge  (x_4 \vee x_5 \vee x_6)                   \wedge (x_1 \vee x_2 \vee \bar{x_3})
\end{array}
$$

As this example suggests, finding solutions of even a modest 3-SAT instance can become difficult quite easily (in this case, $P$ has only one solution: $x_1 = 1, x_2 = 1, x_3 = 0, x_4 = 1, x_5 = 0, x_6 = 0$.)

3-SAT is an NP-complete problem \cite{garey79,sipser05}, as opposed to 2-SAT which can be efficiently solved using a classical computer. Consequently, studying the properties of 3-SAT is
an important area of research, not only because a polynomial-time solution to 3-SAT would imply {\bf P = NP},
but also because 3-SAT (due to its polynomial equivalence with K-SAT) may be used to model problems and procedures
in theoretical computer science \cite{acharyya2007} as well as in several areas of applied computer science and
engineering like artificial intelligence \cite{gent99,mezard02}.

\begin{figure*}[p]
\begin{center}
 \includegraphics[width=0.9\textwidth]{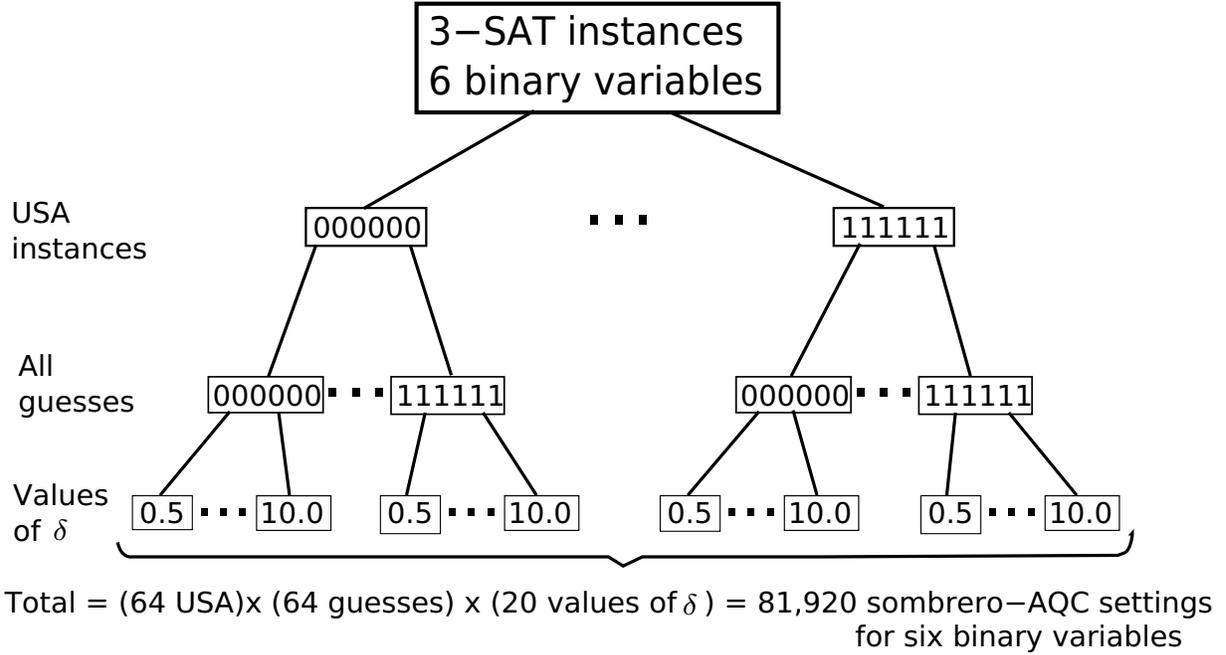}
\end{center}
\caption{\label{fig:calculations-tree}
Scheme for 6 binary variables SAQC calculations.
We generated $2^6$ 3-SAT unique satisfying assignment (USA) instances (first branching), each having as its only solution one of the $2^6$ possible assignments.
All $2^6$ instances have a different state as solution, i.e. there is no chance for repeated instances.
For each instance, we computed minimum-gap values associated with all
possible settings of SAQC (Eq.~\ref{eq:sombrero-aqc}) of
all possible guesses (second branching), using 20 different values
of $\delta \in \{0.5,1.0,\ldots,10.0\}$ 
(third branching). The same scheme
was applied to 7 binary variable 3-SAT USA instances (not shown) for a total of
$(128 \text{USA}) \times (128 \text{guesses}) \times (20 \text{values of } \delta) = 327,680$ SAQC settings.}
\end{figure*}

For the purpose of simplifying our discussion, and without loss of generality, we randomly generated 3-SAT instances
with a unique satisfying assignment (USA) and their number of clauses to number of variables ratio $\alpha \approx 4.26$.
This value of $\alpha$ corresponds to the phase transition region where hard-to-satisfy instances are expected to be found
\cite{achlioptas05,mezard05}. For completeness and to avoid any kind of bias in selecting this pool of instances,
we selected $2^n$ different instances for every $n$ variable case studied. More precisely, to exhaustively study
the impact of different initial guesses with respect to unique solutions
in the behavior of SAQC, we considered all 64 possible initial Hamiltonians $\hat{H}_i$ (using Eq. \ref{h-initial})
for each one of the 64 randomly generated 6-variable USA instances.
Similarly, we built 128 initial Hamiltonians for each 7-variable instance, one per possible initial guess (see Fig. \ref{fig:calculations-tree}). The instances were selected in such a way that their solutions had not only a USA, but also that there was no two instances with the same solution.

The generation of our USA 3-SAT instances took several steps, being the first one using the SAT instance generator developed by \cite{watanabe}.
Unfortunately, as this generator does not warranty the production of 3-SAT instances with unique solutions, we took all generated 3-SAT instances
and determined, by exhaustive bitstring substitution, whether such instances were USA or not. We iterated this process until we computed all
6-variable and 7-variable 3-SAT instances we needed.

The number of qubits used in our simulations is smaller than state-of-the-art calculations for AQC, such as those carried out by quantum Monte Carlo \cite{young08,farhi_quantum_2009}. Here we trade off carrying out few large calculations on many qubits for carrying out many calculations on fewer qubits. We wish to answer the question: What is the the impact of the initial guess in the spectral properties of the time-dependent Hamiltonian? To answer this, we explored the space of initial guesses in an exhaustive manner for the case of
6-variable and 7-variable SAT instances. We run a total of 81,920 and 327,680 for 6- and 7- variables respectively
(see Fig.~\ref{fig:calculations-tree}). As shown in the same figure, we also numerically explored the importance of the strength of the transverse external field for the performance of the algorithm.

Final Hamiltonians $\hat{H}_f$ are instance-dependent, i.e. the structure of each final Hamiltonian
depends on the particular structure (conjunction of clauses) of each 3-SAT USA instance. Our final Hamiltonians  $\hat{H}_f$ comply with the property that it must encode, in its ground state, the solution to
the particular 3-SAT USA instance it was designed for  \cite{Farhi2000,znidaric2005a}. The design of the final Hamiltonian involves an intermediate step,
where a classical cost or energy function is constructed for the particular instance of interest. Once this energy function is expressed in terms of binary variables, it can be easily transformed into a quantum Hamiltonian by performing the mapping indicated by Eq.~\ref{operator-qn}, where each classical binary variable, $q_n$, is transformed into a quantum operator, $\hat{q}_n$. The energy function, $H_f$, associated with the final Hamiltonian, $\hat{H}_f$, can be constructed as a sum of other energy functions, $h_{C_i}$ which involve only variables associated with one clause at a time,
\begin{equation}\label{eq:H_f}
H_f = \sum_i h_{C_i}
\end{equation}
Each $h_{C_i}$ is designed such that it is equal to 1 if clause $C_i$ is unsatisfied and 0 if the clause is satisfied. Notice that the functions $h_{C_i}$ contribute to the count of unsatisfied clauses which defines the spectrum of possible values for $H_f$, with $H_f=0$ when all clauses are satisfied.

Formally, suppose $A=\{x_1, x_2, \ldots, x_n, \bar{x}_1, \bar{x}_2, \ldots, \bar{x}_n \}$ is a set of $n$ binary variables and their corresponding negations, $P$ is a 3-SAT USA instance given by $P = \bigwedge_i C_i$, and each $C_i$ is a disjuction of three elements of $A$, i.e. $C_i = a_\alpha \vee a_\beta \vee a_\gamma$ with
$a_\alpha,a_\beta,a_\gamma \in A$ and indices $\alpha,\beta,\gamma$ are natural numbers, not necessarily consecutive.
Finally, let $B=z_1 z_2 \ldots z_n$ be a set of $n$ bits to be substituted in instance $P$.
Then, $h_{C_i}$ is given by

$$h_{C_i}=
\begin{cases}
0, & \text{if  substitution of } B=z_1 z_2 \ldots z_n \cr
& \text{ in }  a_\alpha \vee a_\beta \vee a_\gamma \text{ makes } C_i = 1\cr

1, & \text{if  substitution of } B=z_1 z_2 \ldots z_n \cr
& \text{ in }  a_\alpha \vee a_\beta \vee a_\gamma \text{ makes } C_i = 0.
\end{cases}
$$

To construct such a function for any arbitrary clause $C_i = a_\alpha \vee a_\beta \vee a_\gamma$, it is useful to note that the only assignment for which $C_i = 0$ is when $a_\alpha =0$, $a_\beta = 0$, and $a_\gamma = 0$. Therefore, the function $h_{C_i}$ by construction, should be 1 when $a_\alpha =0$, $a_\beta = 0$, and $a_\gamma = 0$, and 0 otherwise. It can be easily checked that
 \begin{equation}\label{eq:hclauses}
 h_{C_i} = (1-a_\alpha)(1- a_\beta)(1-a_\gamma),
 \end{equation}
 equals 1 when $a_\alpha =0$, $a_\beta = 0$, and $a_\gamma = 0$ and 0 otherwise. Recall that each $a_\beta$ represents a literal and therefore it could be representing the negation of a variable. One can always use the identity $\bar{x}_i = 1-x_i$ to eliminate any $\bar{x}_i$, and obtain both $h_{C_i}$ and $H_f$ in terms of the $x_i$.

Consider for example the construction of the energy function $h_{C}$ required for clause
$C=\bar{x}_\alpha \vee \bar{x}_\beta \vee \bar{x}_\gamma$, i.e. $C$ is a conjunction of three negated binary variables.
In this case, $C$ is satisfied by all possible 3-bit bitstrings except for $111$ and, according to Eq.~\ref{eq:hclauses}, the energy function assumes the form $h_{C_i}= (1- \bar{x}_\alpha)(1- \bar{x}_\beta)(1- \bar{x}_\gamma) = x_\alpha x_\beta x_\gamma$.

As a last example consider a clause of the form $C=x_\alpha \vee x_\beta \vee x_\gamma$,
where $C$ is a conjunction of three non-negated binary variables taken from set $A$. It is clear that $C$ will be satisfied by all possible
3-bit bitstrings except for $000$. According to Eq.~\ref{eq:hclauses}, the energy function for this clause $C$ is given by $h_{C} = (1-x_\alpha)(1- x_\beta)(1-x_\gamma) =1-x_\alpha - x_\beta - x_\gamma + x_\alpha e_\beta  + x_\beta x_\gamma + x_\alpha x_\gamma - x_\alpha x_\beta x_\gamma$.
From the first equality of the previous equation, one can easily check that $h_C=0$ for all possible 3-bit combinations except for $h_C(000)=1$, as expected.

Once we have the final expression for the final classical energy function of Eq.~\ref{eq:H_f}, the final Hamiltonians $\hat{H}_f$ can be obtained using the mapping of Eq.~\ref{operator-qn}, which relates the classical binary variables with quantum operators. Since we selected only USA instances for our study, each $\hat{H}_{f}$ has a non-degenerate ground state encoding the unique solution of one of our 3-SAT instances with corresponding ground eigenvalue equal to zero. The final Hamiltonians are the same for both strategies, CAQC and SAQC.

\begin{figure*}
\begin{center}
 \includegraphics[width=0.9\textwidth]{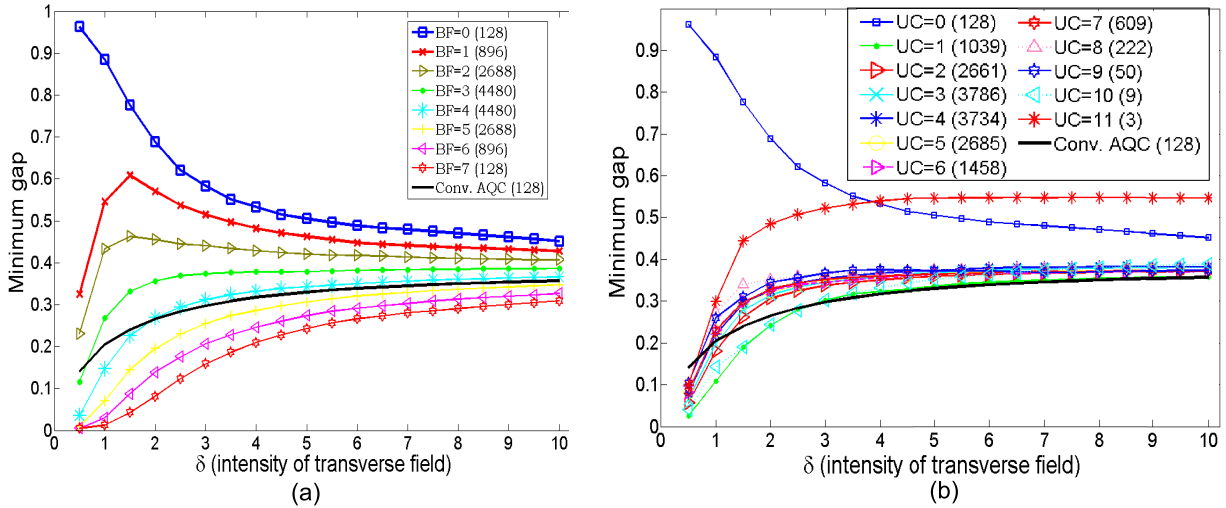}
\end{center}
\caption{\label{fig:median-flips-energies}(Color online)
Summary of the 327,680 calculations for 7 variable 3-SAT instances 
of minimum-gap median values as a function of the transverse field intensity within
groups sorted by (a) number of bit flips, BF, representing the Hamming distance between the initial guess and the solution and (b) number of unsatisfied clauses (UC). Plots include the CAQC (see Eq.~\ref{eq:conventional-aqc}) results for the same 128 different unique-satisfying assignment 3-SAT instances randomly generated for the case of 7 variables. Values in parentheses correspond to numbers of data points which contributed to the value of the median plotted in each curve.
Results for 6 variable instances (not shown) are similar to the ones shown for 7 variable instances.}
\end{figure*}

The numerical results on the dependence of the minimum-gap value, $g_{min}$,
as a function of $\delta$ are shown in Fig.~\ref{fig:median-flips-energies}. Curves 
were computed by taking the median
of all bit strings that fulfilled the criteria specified in the legend boxes; namely either
to produce $i$ unsatisfied clauses (UC=$i$) when substituting the initial guess bit string in its
corresponding instance, or to be $j$ bit flips away from the solution (BF=$j$). BF represents the well known Hamming distance
between the solution and the initial guess state. We focused on UC and BF because, in principle, the notion of closeness of an initial guess
to the actual solution may be defined with either parameter.
Data corresponding to a fixed value of $\delta$ is a statistical representation
(median) of typical $g_{min}$ values that would be expected for hard
3-SAT instances if the guess state belonged to a definite number of UCs or BFs under
an experimental setup using SAQC. Such curves are compared with the minimum gap
expected for CAQC.

\subsection{Effects of the variation in the transverse field intensity on the minimum energy gap}

The dependence of  $g_{min}$ values as a function of the transverse field intensity $\delta$
leaves open some important questions regarding the efficiency of adiabatic quantum algorithms,
whether CAQC or SAQC. For example, what is the optimum value of $\delta$ which minimizes
the running time of an adiabatic algorithm? How transferable is this optimum $\delta$ value among computational problems? Although
we do not intend to do a thorough study of this question in this paper, we would like to give
some insight into this question and provide a qualitative discussion of what kind of results might be expected.

Following closely the notation from Farhi {\it et al} \cite{Farhi2000}, consider $H(t)=
\tilde{H} (t/ \tau) = \tilde{H}(s)$, with instantaneous values of
$\tilde{H}(s)$ defined by
\begin{equation}
 \tilde{H}(s)\ket{E_{l}(s)}=E_{l}(s)\ket{E_{l}(s)}
\end{equation}
with
\begin{equation}
 E_{0}(s) \le E_{1}(s) \le \cdots \le E_{2^N-1}(s)
\end{equation}
where $2^N$ is the dimension of the Hilbert space, and $N$ the number of qubits or equivalently the number of binary variables in the SAT instance. According to the adiabatic theorem, if the gap between the two lowest levels, $E_{1}(s) - E_{0}(s)$, is greater than zero for all $0 \le s \le 1$, and taking,
\begin{equation}\label{eq:runtime}
 \tau \gg \frac{\mathcal{E}}{g_{min}^{2}}
\end{equation}
with the minimum gap, $g_{min}$, defined by,
\begin{equation}
 g_{min}= \min_{0 \le s \le 1} (E_{1}(s) - E_{0}(s)),
\end{equation}
and $\mathcal{E}$ given by,
\begin{equation}
 \mathcal{E} = \max_{0 \le s \le 1} \abs{ \braket{E_{1}(s)}{\frac{d\tilde{H}}{ds} \vert E_{0}(s)} },
\end{equation}
then we can make the normed overlap
\begin{equation}
 \vert \braket{E_{0}(s=1)}{\psi(\tau)} \vert
\end{equation}
arbitrarily close to 1. In other words, the existence of a nonzero gap guarantees that $\ket{\psi (t)}$ remains very close to the ground state of $H(t)$ for all $0 \le t \le \tau$, if $\tau$ is sufficiently large.

Even though we are aware of the new and more stringent conditions for adiabaticity \cite{marzlin_inconsistency_2004,tong_sufficiency_2007,wei2007,zhao2008,amin_inconsistency_2008,mackenzie_validity_2007,ambainis_elementary_2004,jansen_bounds_2007,wu_adiabatic_2007,chen_invariant_2007}
and that Eq.~\ref{eq:runtime} is just one of the inequalities to guarantee adiabatic evolution (though there is still lack of a sufficient
and necessary condition according to Ref.~\cite{du2008}), we will base our discussion on Eq.~\ref{eq:runtime} to illustrate that there is nothing anomalous
in employing the additional Hamiltonian term in the full time-dependent Hamiltonian for SAQC. As well as in CAQC, the algorithmic complexity relies
again in avoiding an exponentially narrowing of $g_{min}$. Along the way we find an important observation about the scaling of the running time as
a function of the parameter $\delta$.

Let us first determine an upper bound for $\mathcal{E}_{SAQC}$ in Eq.~\ref{eq:runtime}. Consider the Hamiltonian in Eq.~\ref{eq:sombrero-aqc}
with hat$(s)=3 s (1-s)$ since this was the functional form used for our numerical calculations. We already discussed, at the end of
Sec.~\ref{subsec:h-init}, that the spectrum of $H_{i}$ is contained in $\{0,1,\cdots,N\}$ and, similarly, it can be easily shown
that the spectrum of $\hat{H}_{driving}$ (see Eq.~\ref{eq:hamilt_perturbation}) is contained in $\{0,1,\cdots,\delta N \}$.
On the other hand, the spectrum of the final Hamiltonian, $\hat{H}_f$, is instance dependent, and its construction guarantees
that the maximum eigenvalue would be $M$ which denotes the total number of clauses. This eigenvalue $M$ would only appear in case
we had an assignment which violates all of the clauses. Using these spectra upper bounds, we can establish an upper bound for
$\mathcal{E}_{SAQC}$ in Eq.~\ref{eq:runtime}, i.e,


\begin{align}
\mathcal{E}_{SAQC} & = \max_{0 \le s \le 1} \bigabs{\braket{E_{1}(s)}{\frac{d\tilde{H}}{ds} \vert E_{0}(s)}} \nonumber\\
 &= \max_{0 \le s \le 1} \bigabs{\braket{E_{1}(s)}{\hat{H}_f-\hat{H}_i+ 3 \delta \hat{H}_{driving} \nonumber\\
 & \quad \quad - 6 \delta s \hat{H}_{driving}   \vert E_{0}(s)}}\nonumber\\
 & \le \max \big( \bigabs{\braket{E_{1}(s)}{\hat{H}_f \vert E_{0}(s)}}+\bigabs{\braket{E_{1}(s)}{\hat{H}_i \vert E_{0}(s)}}\nonumber\\ &+ 3 \bigabs{\delta \braket{E_{1}(s)}{\hat{H}_{driving}\vert E_{0}(s)}} \nonumber\\
 &+ 6 \bigabs{\delta  \braket{E_{1}(s)}{\hat{H}_{driving}\vert E_{0}(s)}}\big) \nonumber\\ &\le M+N+3\abs{\delta} N+6\abs{\delta} N = N(\alpha +1+ 9\abs{\delta})
\end{align}
Where we have used the triangle and Schwartz inequality and also the fact that $M = \alpha N$, with $\alpha$ close to 4.26
in this particular study. We can see that in the worst case scenario, $\mathcal{E}_{SAQC}$ scales linearly with the number
of variables $N$, and linearly with the intensity of the magnetic field, $\mathcal{E}_{SAQC}$ = $O(\abs{\delta}N)$. A similar analysis
gives also that $\mathcal{E}_{CAQC}$ = $O(\abs{\delta}N)$, and therefore, we showed that for SAQC, not surprisingly, the algorithmic complexity also
relies on the scaling of $g_{min}$.

An interesting observation arise by analyzing the linear scaling of $\mathcal{E}$ with $\delta$ and using the numerical
results for the dependence of the typical minimum gap values as a function of $\delta$.
There seem to be at least two distinguishable regimes for the dependence of $g_{min}$ on $\delta$ for both CAQC and SAQC (Fig.~\ref{fig:median-flips-energies}).
For relatively small values of $\delta \in [0.5,1.5]$, $g_{min}$ scales approximately
linearly with $\delta$ and therefore $g^2_{min} \sim \delta^2$. Since the running time
is given by Eq.~\ref{eq:runtime}, and $\mathcal{E} \sim \delta$, the running time $\tau$ decays inversely proportional to
$\delta$ within this linear regime. However, for large values of $\delta$, in the \lq stationary' regime where $g_{min}$ is
almost constant, increasing field intensity through $\delta$ would make both algorithms
less efficient as running time $\tau$ would increase roughly linearly with $\delta$.

Both CAQC and SAQC would benefit from an increase in the transverse field for small values of $\delta$,
but notice that $g_{min}$ values for SAQC are more sensitive to $\delta$,
and soon become better on average than those for CAQC (Fig.~\ref{fig:median-flips-energies}). According to the previous discussion about
running time as a function of $\delta$, it would be ideal to choose $\delta$ near the end of the linear regime; in our calculations,
$\delta$ somewhere in the interval (1,2). Further studies concerning the optimum value of $\delta$ as a function of the number of binary variables are needed, but we chose $\delta = 1.5$ for our analysis on the performance in SAQC and CAQC described in the following section.

\subsection{Performance comparison between SAQC and CAQC}

\begin{figure*}[p]
\begin{center}
 \includegraphics[width=0.9\textwidth]{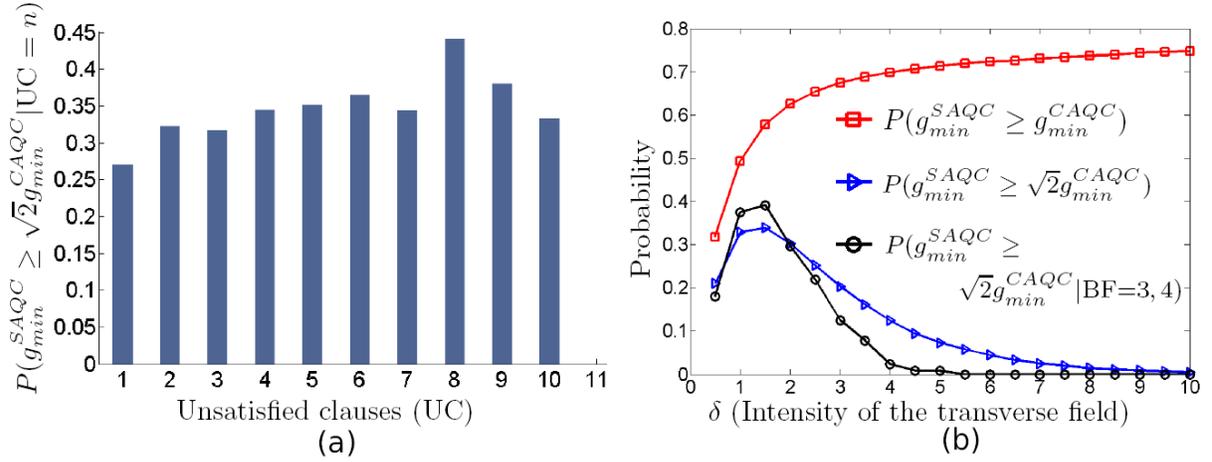}
\end{center}
\caption{\label{fig:probability-plots}(Color online) (a) $P(g_{min}^{SAQC} \ge \sqrt{2} g_{min}^{CAQC} | UC = n)$
is the conditional probability of choosing a state with UC$=n$ and a SAQC minimum-gap large enough so that the performance of the SAQC is \textit{significantly better} (at least twice faster) than the CAQC. The results were obtained for $\delta = 1.5$. Panel (a) shows that there is no correlation between the number of violated clauses and the $g_{min}$ of the SAQC algorithm, for the hard-to-satisfy instances randomly chosen for this numerical study. Panel (b) shows the probability of choosing an initial state at random and satisfying the condition specified in the legend, for different values of the transverse field intensity, $0.5 \le \delta \le 10$. The conditional probability $P(g_{min}^{SAQC} \ge \sqrt{2} g_{min}^{CAQC} | BF = 3,4)$ (triangles) aims to predict the performance of the SAQC algorithm in the case of large number $n$ of qubits. In this limit, an initial state chosen at random will have with high probability a Hamming distance $BF \sim n/2$, given that they are binomially distributed, i.e., the number of $n$ bit strings with BF $= m$ is equal to $\binom{n}{m}$}
\end{figure*}

The data sorted with respect to BFs and UCs shows an increase of the minimum gap, $g_{min}$, as the Hamming distance from the initial guess to the solution decreases; the trend for UC is less apparent (see Fig.~\ref{fig:median-flips-energies}). Computing the number of UCs produced by
a given initial guess can be done in polynomial time on a classical computer. Unfortunately there is no way to determine \textit{a priori} how many bit flips the guess is from the solution, as that requires knowledge of the solution itself. Additionally, Fig.~\ref{fig:probability-plots}(a) shows that the SAQC implementation,
using $\hat{H}_{driving}$ as defined in Eq.~\ref{eq:hamilt_perturbation}, does not necessarily favor states with low values of UC, but rather gives a homogenously distributed success probability between 25-45\%, for $\delta=1.5$. This is in accordance to the observation that 
solving 3-SAT hard instances is not necessarily guided by minimizing the number of UCs \cite{kautz07}. Given the above scenario, we analyzed the likelihood of better performance by choosing initial guesses at random.

In the following discussion, we use the term \textit{significantly better} initial guess
to mean an initial condition that leads a SAQC algorithm to be at least twice as fast as CAQC,
i.e. running times for CAQC, $\tau_{CAQC}$, and SAQC, $\tau_{SAQC}$, are such that
$\tau_{CAQC}\geq 2 \, \tau_{SAQC} $ or, equivalently,
$g_{min}^{SAQC} \geq \sqrt{2} \, g_{min}^{CAQC}$,
assuming $\mathcal{E}_{CAQC} = \mathcal{E}_{SAQC}$.

For $\delta=1.5$, choosing an initial state at random yields a probability greater than $50\%$ of having
$g_{min}^{SAQC} \geq g_{min}^{CAQC}$ (squares) as shown in Fig.~\ref{fig:probability-plots}(b). Moreover, the probability of \textit{significantly better} performance, i.e. $\tau_{CAQC} \geq 2 \, \tau_{SAQC}$ is $\approx 35 \%$ (triangles). With the intention of predicting the performance of the SAQC protocol in the limit of large $n$, the third curve (circles) was produced using the following rationale:
for USA instances, the number of bit configurations with a given value of $BF=m$ follows a binomial distribution $\binom{n}{m}$.
In the limit of large $n$, the likelihood of choosing a state in the central region of the binomial distribution
is the highest. This observation led us to concentrate on the performance of the most populated instance subsets,
those that correspond to BF $=3,4$ for 7 variables. Here, the probability of significantly better performance is close to $40\%$.

\begin{figure*}
 \begin{center}
  \includegraphics[width=0.9\textwidth]{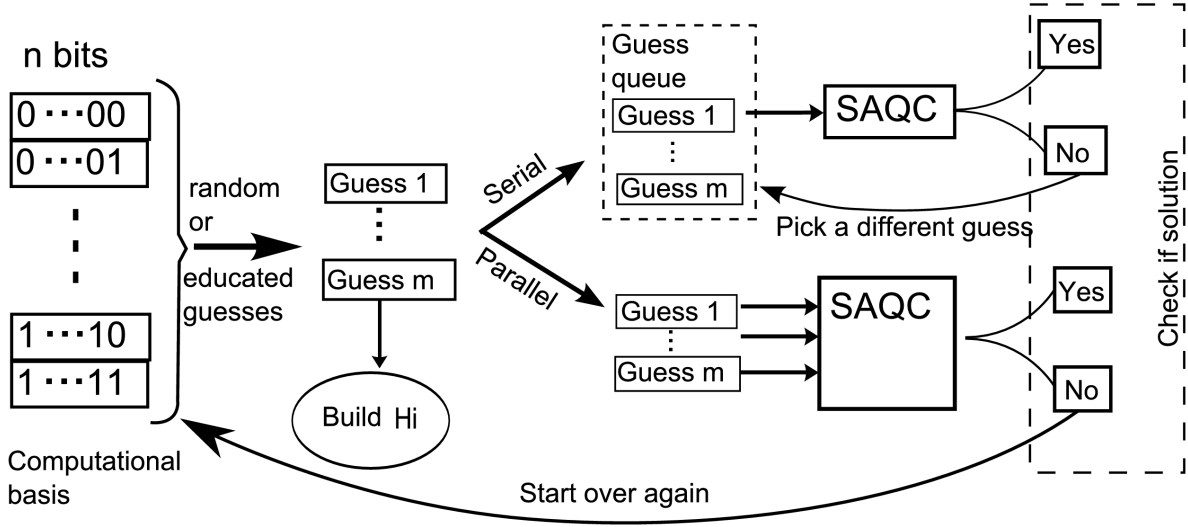}
 \end{center}
 \caption{\label{fig:sombrero-AQC-algorithm}(Color online) Implementation of an SAQC algorithm either in parallel or in serial. The algorithm begins by choosing a state from the computational basis. For each chosen initial state, an initial Hamiltonian is prepared according to Sec.~\ref{subsec:h-init}. Next, choose an ideal time assuming a CAQC protocol will be run, and use that as a reference to run the SAQC protocol twice as fast. If only one AQC computer is available, one can still use the probabilistic speed up obtained in SAQC to run for example two adiabatic protocols instead of one, in serial mode. Once the first SAQC calculation is finished, one can efficiently check whether or not the result is a solution. In case that it is not a solution, one can submit an additional calculation, either randomly selecting another initial guess state or using the measured excited state. We call this latter option, ``quantum heuristics" since the outcome of the near-adiabatic quantum evolution is used to refine the initial guess state for further experiments. In the case of having several adiabatic quantum computers at hand, one can do the same initial procedure of selecting guesses, but now submitting a different guess to a different node and running on each node twice as fast.}
 \end{figure*}

Finally, we propose an algorithm based on SAQC. An initial guess is chosen either at random or by applying expert-domain knowledge and then encoded into the initial
ground state of $\hat{H}_i$ (Eq.~\ref{h-initial}). An adiabatic passage based on SAQC is then performed either in serial
or in parallel, depending on the availability of quantum hardware resources (see Fig.~\ref{fig:sombrero-AQC-algorithm}.). As an example of the potential usefulness of our algorithm,
recall from Fig.~\ref{fig:probability-plots} that the probability of \textit{significantly better} performance using SAQC is $39\%$ for $\delta=1.5$.

One way to employ the probabilities we obtained from our numerical simulations in a more concrete scenario is:
suppose one is assigned the task of using an adiabatic quantum computer and assume one uses Eq.~\ref{eq:runtime} or any
of the more stringent conditions for adiabaticity \cite{marzlin_inconsistency_2004,tong_sufficiency_2007,wei2007,zhao2008,amin_inconsistency_2008,mackenzie_validity_2007,ambainis_elementary_2004,jansen_bounds_2007,wu_adiabatic_2007,chen_invariant_2007} to estimate for how long
one may need to run an algorithm under the CAQC paradigm.  Moreover, suppose that this estimated running time required
to remain in  the ground state with a high success probability is $\tau_{CAQC}=$ 2 days. Using the numerical results presented in this paper,
one can opt for performing  the same task using SAQC as follows: suppose we have absolutely no information about
the problem as it was in the case of this numerical study with random 3-SAT USA instances. Instead of running
the CAQC algorithm for 2 days, pick a guess state at random from all the possible assignments and then use it
as initial state for SAQC and run the algorithm with $\tau_{SAQC}=$ 1 day.
According to Fig.~\ref{fig:probability-plots}, the probability of having picked a state whose performance is as good as CAQC is 39\%,
for $\delta= 1.5$. If after measurement at the end of the first day the result is not a solution,
we still can pick another state at random and let it run for one more day. By now the probability of having picked
a state with the same performance as the CAQC in the two trials equals 63\%. Note that in this simple probability calculations we are not
taking into account the fact that even in the case where the state selected for the first run was not one of the \lq ideal' ones
(states corresponding to the 39 \% of guesses for the results presented in Fig.~\ref{fig:probability-plots}), we still have
a very good chance that the \lq non-ideal' state still delivers a right answer after the first measurement. This probabability
will depend of course on how close is the chosen state from the set of \lq ideal' ones.

Consequently, the execution of two SAQC algorithms in serial would take at most as much time as
the execution of only one CAQC algorithm. By allowing us to choose two guesses to run in the same time as one case in CAQC,
the probability of choosing a significantly better initial guess in these two SAQC executions increases from 39\% to 63\%.
Furthermore, even when no significantly better initial guess is chosen and the process is not guaranteed to be fully adiabatic,
there is still some probability that we measure the correct solution at the end of both executions.

\section{Conclusions}\label{sec:conclusions}
In summary, we propose an algorithmic strategy which incorporates heuristics in adiabatic quantum computation. In particular, we study one of the most basic heuristic strategies consisting of initializing the computation with a desired initial state chosen by physical intuition, an educated guess, classical preprocessing of the problem, and/or by randomly choosing one of the possible assignments. This method allows to bridge powerful classical techniques such as heuristic optimization routines and/or mean-field calculations to obtain approximated solutions of the problem and use them as initial guess states for SAQC. The strategy presented allows for a parallel and/or a concatenated scheme. The parallel setup might be helpful if several adiabatic quantum platforms are available in which several guesses can be run simultaneously, one guess to run in each one of the adiabatic processing units. The idea of the concatenated scheme is to restart a failed adiabatic evolution with the measured excited state. Assuming a near-adiabatic trajectory, the measured excited state can be taken as a refined guess from which the quantum evolution is restarted. Notice this is not possible in the conventional approach which would restart the evolution from the full superposition, attributing equal probability to all the states, therefore ``erasing" the information gained in the previous near-adiabatic evolution. Neither of the two features proposed above are possible in any of the different adiabatic quantum computation proposals to date. In addition, all the modifications proposed related to different adiabatic paths, auxiliary perturbations \cite{farhi02} can also be explored in the context of SAQC, which is also suitable to study quantum problems
\cite{kempe_complexity_2006,bravyi_efficient_2006,oliveira_complexity_2008,aharonov07,mizel07}.

The numerical study performed in this paper is a proof-of-principle to explore the importance and consequences of starting the adiabatic evolution with a guess state and to illustrate that incorporating this kind of heuristics in AQC is possible and might be advantageous. Our numerical simulations show that starting the adiabatic evolution with a guess state which has a zero-overlap with the solution, $\braket{\phi_{guess}}{\phi_{solution}} = 0$, is not a big concern. On the contrary, even when there is no hope to make an educated guess and selection of the initial state at random is the only available alternative, we obtain that approximately 40\% of states might allow running the quantum algorithm at least twice as fast when compared to CAQC. This possibility of running the algorithm for shorter times
but with several trials brings also additional advantages of getting the right answer in any intermediate measure. Moreover, these shorter
runs would be less affected by decoherence effects.

Even though the procedure used for the performance comparison, CAQC \textit{vs.} SAQC, is ``reasonable" since it is based on the widely-used minimum gap criteria and its connection with the algorithm run-time (see Eq.~\ref{eq:runtime}), we are aware of the limitations of this analysis    ~\cite{marzlin_inconsistency_2004,tong_sufficiency_2007,wei2007,zhao2008,amin_inconsistency_2008,mackenzie_validity_2007,ambainis_elementary_2004,jansen_bounds_2007,wu_adiabatic_2007,chen_invariant_2007}. We want to stress that the present numerical results are only encouraging indicators that heuristics in AQC might be a valuable algorithmic strategy for AQC, given the strong dependence of the value of the minimum gap as a function of the initial guess chosen. It is not our purpose to claim superiority of SAQC but to introduce the approach and the motivation behind it. For a more rigorous comparison of both schemes, CAQC and SAQC, we suggest numerical experiments which are not meant to be exhaustive but preferably involving larger size instances, and to explore different problems other than random 3-SAT. For example, we are interested in performing these studies for relevant instances of our recently developed AQC proposal for protein folding \cite{perdomo08} and for adiabatic preparation of molecular ground states \cite{Aspuru-Guzik2005} where we expect the mean-field (Hartree-Fock) solution to be a better guess than a full superposition. Numerical propagation of the time-dependent Schr\"{o}dinger equation, instead of inspection of the minimum gap after the Hamiltonian diagonalization, for these cases will provide a realistic simulation of the quantum computation.

 Open questions to be explored further is the connections between SAQC,
 quantum phase transitions \cite{schutzhold06,latorre_adiabatic_2004}, entanglement \cite{ors_universality_2004}
 and the effect of local minima \cite{amin2008}. The performance of the adiabatic algorithm in the limit of large $n$ is still an open question \cite{farhi_quantum_2009,young_2009} which needs to be explored in the context of SAQC. It is not obvious that the same observations and conclusions/observations mentioned above for CAQC will hold for SAQC as well. For example, we think that SAQC might be a strategy to avoid the local minima traps described in Ref.~\cite{amin2008}. From the study of the spectral properties of the random 3-SAT instances studied, we found that some initial states have a considerably larger gap and while others show a considerably smaller gap when compared with CAQC. Since the initial state in CAQC is a full superposition of all the possible states or solutions, including the ones with a large gap and others with small gap, we conjecture here that having a full superposition will not necessarily be the best choice, given that the presence of the states with small gaps could slow down the quantum evolution. An analysis beyond the gap criteria would be needed to test this conjecture. Solving the time-dependent Schr\"{o}dinger equation for the entire evolution is the most straightforward, yet numerically-challenging approach. In SAQC, the probability of obtaining these trap states can be avoided and even in the case of a failed evolution, the concatenated scheme may help to restart using a better guess. In contrast, in CAQC, the full superposition including states with considerably small gap might result in a bottleneck for the dynamics towards a successful computation. Further studies need to be done to verify this is indeed the case.

\begin{acknowledgements}
  All authors thank E. Farhi and S. Gutmann for useful discussions the led to Fig.~\ref{fig:probability-plots} and related analysis. S.V.A. gratefully acknowledges
  P. Grasa and R. Rueda for their support, and Tecnol\'{o}gico de Monterrey
  Campus Estado de M\'{e}xico for funding. A.P. and A.A.G gratefully acknowledge
  funding from D-Wave Systems and Harvard's Institute for Quantum Science and
  Engineering. A.A.G. thanks the Army Research Office under contract W911NF-07-0304.
\end{acknowledgements}




\end{document}